\documentstyle[prd,aps,floats,times,epsfig]{revtex}
\draft
\def\be{\begin{equation}}
\def\ee{\end{equation}}

\begin{document}
 {}~\hfill\vbox{\hbox{gr-qc/0103098}\hbox{MRI-P-010303}}
\vskip 4mm
\centerline{\large{\bf  Cosmology in scalar tensor theory and
asymptotically de-Sitter Universe}}
\vskip 3mm
\centerline{\bf A.A.Sen\footnote{anjan@mri.ernet.in} and
  S.Sen\footnote{somasri@mri.ernet.in}}
\vskip 3mm
\centerline{Harish-Chandra Research Institute, Chhatnag Road,
Jhusi. Allahabad 211
019 India}
\date{{\today}}
\vskip 2mm
\begin{abstract}
We have investigated the cosmological scenarios with a  four dimensional effective
action which is  connected with multidimensional, supergravity and string
theories. The solution for the scale factor is
such that initially universe undergoes a decelerated expansion but in late
times it enters into the accelerated expansion phase. Infact, it
asymptotically becomes a de-Sitter universe. The dilaton field in our model
is a decreasing function of time and it becomes a constant in late time
resulting the exit from the scalar tensor theory to the standard Einstein's gravity. Also
the dilaton field results the existence of a positive cosmological constant
in late times.
\end{abstract}
\pacs{PACS Number(s): 04.20Jb, 98.80Hw}
\vskip 2mm
The observed location of the first acoustic peak of CMB temperature
fluctuation corroborated by the latest BOOMERANG and MAXIMA data \cite{cmb}
favours a spatially flat matter dominated universe whose energy density is
dominated by some {\it missing energy component}. On the other hand, recent
measurements of the luminosity-redshift relations observed for a number of
newly discovered type Ia supernovae indicate that at present the universe is
expanding in an accelerated manner suggesting that this {\it missing energy}
has a negative pressure\cite{SN}. The first and obvious choice for this missing
component is the vacuum energy or cosmological constant $\Lambda$. However
this possibility of $\Lambda$ being the dominant component of the total energy
density of the universe, has problem that the energy scale involved is lower
than normal energy scale of most particle physics model by factor of $\sim
10^{-123}$. The next choice for this missing energy is a dynamical
$\Lambda$ in the form of a scalar field slowly rolling down a considerably
flat potential so that its slowly varying energy density mimics an effective
cosmological constant. This form of missing energy density is called {\it
Quintessence} and this is similar to the inflation field with the difference
that it evolves in a much low energy scale.

There are a number of quintessence models which have been put forward in
recent years and most of them involve a minimally coupled scalar field with
different type of potentials dominating over the kinetic energy of the scalar
field giving rise to an effective negative pressure \cite{Min}. Although all
these models have their own
merits in explaining the missing energy and the accelerated expansion of the
universe, they also have some drawbacks. First of all the minimally coupled
self interacting scalar field models will be ruled out if the observations
predicts that the missing component of the energy density obeys an equation of
state $p=\gamma\rho$ with $\gamma<-1 (\rho\geq 0)$, and this sort of equation
of state is in reasonable agreement with different observations
\cite{eqn}. Also the inequality $dH^{2}(z)/dz\geq 3\Omega_{m0}H_{0}(1+z)^{2}$
should have to be satisfied for minimally  coupled scalar field and its violation
will certainly point towards a theory of non Einstein gravity such as scalar
tensor theories where the scalar field is nonminimally coupled to
gravity. 

Also if one considers a supergravity theory in higher dimensions, 
then gravity is
in general coupled to gauge fields composing the bosonic sectors. The
reduction to four dimensions leads to a nontrivial coupling between the
gravity and scalar fields, called dilaton, some of which come from the
compactifications of the internal dimension. The cosmological scenarios of
these effective models in four dimensions has been extensively 
studied \cite{string}.
In recent years there are also attempts in modelling the missing energy of the
universe and to explain its late time accelerated expansion in perview of
these scalar tensor theories where the scalar field is nonminimally coupled to
gravity sector \cite{nonmin}. Attempts have also been done to study the late 
time accelerated
expansion of the universe in context of original Brans-Dicke (BD)theory 
where the
BD scalar field is massless \cite{BP,ASS}. One of the problems in these models
is that it is difficult to incorporate the decelerating expansion phase of the
universe in these models. Hence these models are always accelerating which
seriously contradicts with the big bang nucleosynthesis and structure formation
scenario of the universe. It was shown by Banerjee and Pavon \cite{BP} that
one can avoid such problem by assuming the BD parameter to be a function of
the BD scalar field $\phi$. Also all these scalar tensor theories
are labelled by a parameter $\omega$
and another serious problem is that in most of
these models, in order to have the accelerated expansion of the universe in
late time together with the right estimates for $\Omega_{m}$ and
$\Omega_{\phi}$, the range of this parameter $\omega$ is too small to match 
 with the solar system experimental bound
$\omega>500$. Bertolami and Martins\cite{nonmin} have obtained the solution
for the accelerated expansion of the universe in BD cosmology with a
$\phi^{2}$ potential with large $|\omega|$. But in their work they have not
considered the positive energy conditions for the matter and the scalar field.
The reasons for these problems may be that in most of these models the simple
power law expansion of the universe is assumed. Hence it may be worthwhile to
study some different expansion behaviour of the universe which can give
rise to late time accelerated expansion of the universe and also can solve the
problems described above.

In the present work, we have used the low energy four dimensional effective action for the
higher dimension theory, together with a matter fluid having a dissipative
pressure over and above its positive equilibrium pressure. The CDM
is in general considered to be a perfect fluid having a zero positive equilibrium
pressure.  However  it has been proposed recently that the CDM must
self interact in order to explain the detailed structure of the galactic halos
\cite{CDM}. This self interaction may create a pressure for the CDM which can
also be negative\cite{chpav}.
Also, as demonstrated in a recent paper by Zimdahl et.al \cite{zim}, one can
have a pressure in the CDM if there exists an interaction which does
not conserve particle numbers, and negative pressure can exist due to particle production out of
gravitational field. In this case, the CDM is not a conventional dissipative
fluid, but a perfect fluid with varying particle number. Zimdahl et.al have
shown that even extremely small particle production rate can cause the
sufficiently negative pressure to violate the strong energy condition. In our
calculations, we have not assumed any particular model for the negative
pressure of the CDM rather we have studied its effect in the late time
expansion of the universe. Unlike the previous works on nonminimally coupled
scalar tensor theories, we have not assumed any particular form for the
potential for the BD scalar field. Instead we have assumed the temporal
dependence of the BD scalar field in such way that one can
recover the Einstein's gravity in present day. This will ensure that there is
no conflict with the present day Solar system experimental results which are
very much consistent with the Einstein's gravity. The solution for the
scale factor of the universe is such that initially the universe undergoes
decelerated expansion which is necessary for the structure formation scenario
in the matter dominated universe, but in late time it enters the accelerated
expanding  phase suggested by the recent Supernova observations. In fact it
has been shown that one can get a de-Sitter expanding universe asymptotically.
The potential we obtain for such solution is a combination of terms like
$\phi^{p}(ln(\phi))^{q}$ with $p$ and $q$ taking different values. This kind
of potential has earlier been studied for inflationary model with
minimally coupled scalar field by Barrow and Parson\cite{BPA}. Terms such as
this also appears in the Coleman-Weinberg potential for new
inflation\cite{new}.  The behaviours of energy
densities of the matter and the  dilaton field are shown to be quite
satisfactory in our model.

The field equations derived from the low energy effective action of the string
theory is given by
\be
G_{\mu\nu} = \frac{T_{\mu\nu}}{\phi}+\frac{\omega}{\phi^2} (\phi_{,\mu} \phi_{,\nu}
-\frac{1}{2}g_{\mu\nu}\phi_{,\alpha}\phi^{,\alpha})
+\frac{1}{\phi}[\phi_{,\mu;\nu} - g_{\mu\nu}\Box{\phi}]
-g_{\mu\nu}\frac{V(\phi)}{2\phi},
\label{fldeqn}
\ee
where $T_{\mu\nu}$ represents the energy momentum tensor of the matter 
field, $\phi$ is the dilaton field and $\omega$ is a dimensionless parameter. We have assumed the matter content of the universe to be composed 
of a fluid represented by the energy momentum tensor 
\be
T_{\mu\nu}=(\rho+P){\it{v}}_\mu{\it{v}}_\nu+P
g_{\mu\nu},\label{emtensor}
\ee
where $\rho$ and $P$ are the energy density and effective pressure of 
the fluid respectively and  ${\it v}_\mu$is the four velocity of the fluid
i.e, ${\it{v}}_\mu{\it{v}}^\mu=-1$. The effective pressure of the fluid 
includes the thermodynamic pressure $p$ and a negative pressure $\pi$, 
which could arise either because of the viscous effect or due to particle 
production, i.e, 
\be
P=p+\pi
\label{pressure}
\ee
The dynamics of the dilaton field is governed by the equation,
\be
\Box\phi=\frac{T}{2\omega+3}+\frac{1}{2\omega+3}\left(\phi\frac{dV(\phi)}
{d\phi}-2V(\phi)\right)\label{waveeqn}
\ee 
where $T$ is the trace of $T_{\mu\nu}$. The background metric is considered to be standard Friedman-Robertson-
Walker one with the 
signature convention $(-,+,+,+)$ and $R$ is the scale factor. 
We restrict ourselves for spatially
flat metric only. We work in Jordan frame. One interesting thing about 
working  in Jordan frame is that the conservation equation holds for 
matter and scalar field separately. Or in a slightly different way, the 
Bianchi Identity along with the wave equation (\ref{waveeqn}) gives 
the matter conservation equation
\be
\dot\rho+3{\dot R\over{R}}(\rho+p+\pi)=0
\label{coneqn}
\ee
Expressing explicitly in terms of the FRW metric the field equations and 
the wave equation respectively appear to be
\be
3{\dot R^2\over{R^2}}+3{\dot
R\over{R}}{\dot\phi\over{\phi}}-{\omega\over{2}}{\dot\phi^2\over{\phi^2}}-
{V\over{2\phi}}={\rho\over{\phi}},
\label{fldeqn1}
\ee
\be
2{\ddot{R}\over{R}}+{\dot R^2\over{R^2}}+{\ddot
{\phi}\over{\phi}}+2{\dot
R\over{R}}{\dot\phi\over{\phi}}+{\omega\over{2}}{\dot\phi^2\over{\phi^2}}-
{V\over{2\phi}}=-{p\over{\phi}}
\label{fldeqn2}
\ee
\be
{\ddot{\phi}}+3{\dot
R\over{R}}{\dot\phi}={\rho-3p\over{2\omega+3}}-{1\over{2\omega+3}}\left[2V-
\phi{dV\over{d\phi}}\right]
\label{wveqn}
\ee

In this situation we are at liberty to make some assumptions as we have more 
unknowns $(R, \phi,\rho,p,\pi,V)$ with lesser numbers of equations to 
determine them. 
Unlike the previous investigations where people have assumed the power law
expansion of the universe together with some suitable form of the potential, we will assume the form for the dilaton
field in such a way so that it stabilises quickly with time resulting the
transition to GR. Subsequently we calculate the form of the potential which
can give rise to this kind of behaviour.

We assume the ansatz
\be
\frac{\dot \phi}{\phi}=-\beta Exp[-\alpha(t-t_0)]
\label{phidot}
\ee
and 
\be 
H=\frac{\dot R}{R}= H_1 Exp[-\alpha(t-t_0)]+H_2
\label{h}
\ee
where $H_1,~H_2,~\alpha,~\beta$ and $t_0$ are all positive constants.
This essentially means that we are working in a scenario where the scale 
factor and the scalar field evolve as 
\be
R=R_0 Exp[H_2 t - \frac{H_1}{\alpha} e^{-\alpha(t-t_0)}]
\label{R}
\ee
and
\be
\phi=\phi_0 Exp[\frac{\beta}{\alpha}e^{-\alpha(t-t_0)}]
\label{phi}
\ee

As is evident from this expression, the dilaton  field is an 
exponentially decreasing function of time and quickly stabilises 
at some constant $\phi_0$ for large time. Hence in late time one could 
recover General Relativity by identifying $1/\phi_0$ with $G_0$, the 
present Newtonian constant. Hence whatever be the choice of $\omega$ 
we have in our model. that really does not contradict the solar system 
bound on $\omega$ as for the present day the theory becomes GR.\\
\begin{figure}
\centering
\leavevmode\epsfysize=7cm \epsfbox{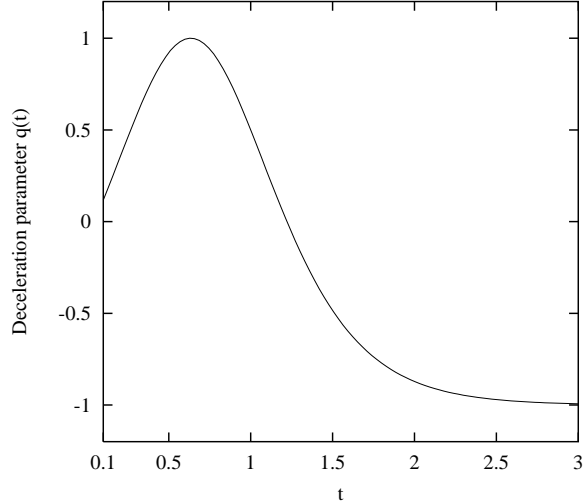}
\caption{The deceleration parameter $q$ {\it vs} time $t$ in units of 10 Gyrs}
\label{fig1}
\end{figure}
The deceleration parameter seems to be an important parameter to specify 
the expansion of the universe. For last several decades the standard 
cosmological models favoured  a presently matter dominated universe 
expanding in a decelerated fashion. The positive value of the 
deceleration parameter was more or less compatible with all 
cosmological tests. But the recent observation seem to favour 
negative value of this parameter to portray an accelerated expansion 
of the universe. With an ansatz of the form given by (\ref{h}), 
the deceleration parameter $q$ is given by

\be
q=-\frac{\dot{H}}{H^2}-1=\frac{\alpha H_1e^{-\alpha(t-t_0)}}
{[H_2+H_1e^{-\alpha(t-t_0)}]^2}-1
\label{q}
\ee
The behaviour of the deceleration parameter $q$ is presented in 
figure 1. For this we have chosen the parameter $t_0$ 
to be $10^{10}$ yrs and also  we have set our initial time to $10^9$ yrs, 
which is approximately the time for first bound structures to be 
formed\cite{LL}.   
The graphical representation of $q$ shows that initially it has a positive value 
suggesting decelerated expansion, but with time it  attains a negative value 
depicting accelerated expansion and finally saturates to $-1$. This 
feature of $q$ quite satisfactorily explains the evolution of the 
universe where the matter dominated universe  decelerates but starts 
accelerating  at late time as suggested by present 
observation\cite{SN}. It is interesting to notice that the universe 
starts accelerating very recently which is quite consistent with 
present estimated age of the universe\cite{Pont}.
Also, the figure shows
that it asymptotically the expansion becomes de-Sitter suggesting the existence
of a positive cosmological constant. 
\begin{figure}
\centering
\leavevmode\epsfysize=7cm \epsfbox{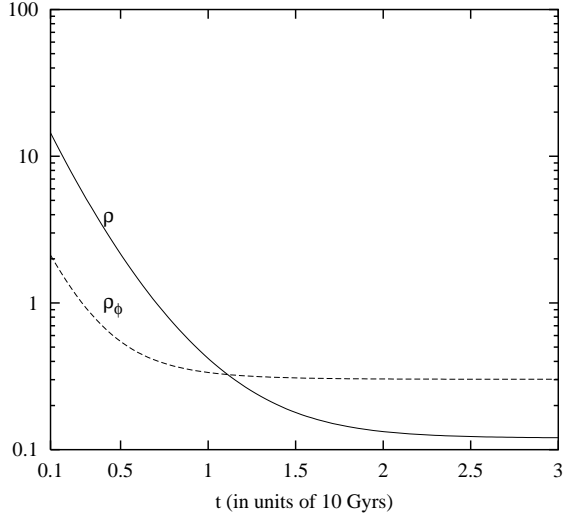}
\caption{Energy densities $\rho$ and $\rho_{\phi}$ {\it vs} time $t$}
\label{fig2}
\end{figure} 
Apart from the 
appropriate form of evolution of the universe, constraints are imposed 
on an acceptable model from the ratio of the energy densities of matter 
and scalar field. In the early times, such as at $10^9$ yrs, the matter 
energy density dominates the total energy density to justify the 
structure formation scenario, whereas at late time the scalar energy 
density plays the predominant role to explain the so called missing 
energy density of the universe.

\begin{figure}[hb]
\centering
\leavevmode\epsfysize=7cm \epsfbox{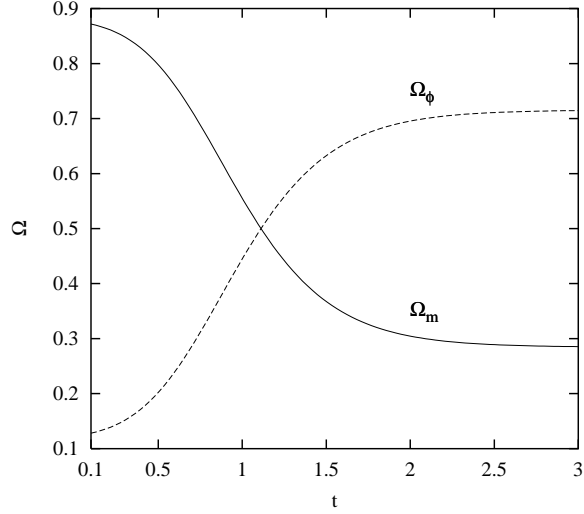}
\caption{Density parameters $\Omega_m$ and $\Omega_{\phi}$ {\it vs} time $t$}
\label{fig3}
\end{figure}

To find whether the present ansatz 
accommodates these constraints we find the energy densities of the matter 
and the scalar field. With this specific kind of evolution of the 
universe [(\ref{h}),(\ref{phi})] the energy density of the matter and 
the scalar field turns out to be respectively
\be 
\rho= \left\{\frac{3H_2}{8}(16H_1-9\beta)e^{-\alpha(t-t_0)}
-3H_1[\beta(1+\omega)+H_1]e^{-2\alpha(t-t_0)}\right\}\phi_0Exp[\frac{\beta}
{\alpha}e^{-\alpha(t-t_0)}]+\rho_0 
\label{ro}
\ee
and 
\be 
\rho_\phi= \left\{3H_2^2+\frac{27}{8}\beta H_2e^{-\alpha(t-t_0)}
+3H_1[\beta(1+\omega)+2H_1]e^{-2\alpha(t-t_0)}\right\}\phi_0Exp[\frac{\beta}
{\alpha}e^{-\alpha(t-t_0)}]-\rho_0.
\label{rop}
\ee
where $\rho_0$ is an integration constant with the choice
\be 
\omega+1=-\frac{(16H_1-3\beta)^2}{8\beta(16H_1-\beta)}{~~~\rm and~~~}
\alpha=8H_2.  
\label{w}
\ee
In order to ensure the positive energy condition for both matter and scalar
field the constraints on the constants are $(16H_1-3\beta)^2>8H_1(16H_1-\beta)$ and $16H_1-9\beta>0$. In figure \ref{fig2} we plot 
the the energy densities
of both the matter and the scalar field. 
The energy scale is plotted in units of $10^{-47}$ GeV$^4$. We 
find that the above constraints are accommodated i.e, the matter energy 
dominates the scalar energy in the early time but it evolves at such a 
rate that at late time the scalar energy density takes over the matter 
density and finally tracks to a constant ratio of the two densities. 
This is a very important behaviour so far as the late time acceleration 
and quintessence scenario are concerned. This tracking behaviour is a 
possible solution to the cosmic coincidence problem of the quintessence 
proposal. An interesting point to note here is that  the energy density 
of matter $\rho^{1/4}$ is $\sim 10^{-3}$ eV at an epoch 1 Gyrs when the 
first bound structures were formed, which agrees quite well with the 
available data\cite{LL}.
On the other hand the density parameters also gives a clear picture about 
the ratio of energy densities of the universe at different epochs. A plot 
of the density parameter of matter $(\Omega_m=\frac{\rho}{3H^2\phi})$ and 
that of scalar field $(\Omega_\phi=\frac{\rho_\phi}{3H^2\phi})$ is shown 
in figure \ref{fig3}. $\Omega_m$ dominates the $\Omega_{tot}$ in the 
early time, but off late 
$\Omega_{\phi}$ becomes the major contributor of $\Omega_{tot}$ and finally 
both $\Omega_m$ and $\Omega_{\phi}$ saturates at values $0.28$ 
and $0.72$ respectively, which match the recent observations 
excellently\cite{SN}. \\

\begin{figure}[hb]
\centering
\leavevmode\epsfysize=7cm \epsfbox{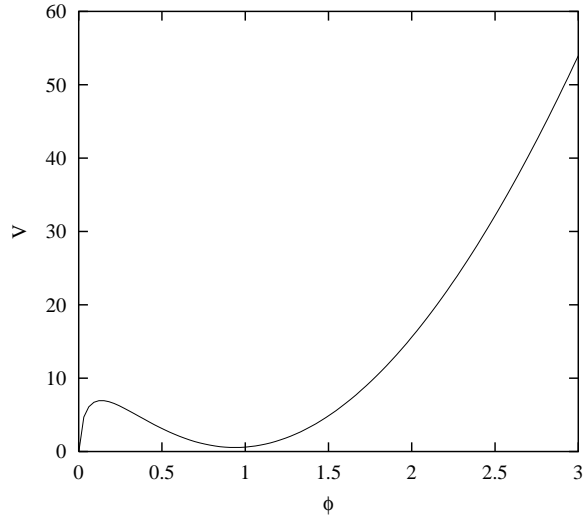}
\caption{Plot of potential}
\label{fig4}
\end{figure}  

The form of potential admissible in this ansatz is given by
\be
V(\phi)=\left(\frac{8H_2}{\beta}\right)^2[3H_1(\beta\omega+2H_1)-
\frac{\omega}{2}\beta^2]\phi\left[\ln\frac{\phi}{\phi_0}\right]^2
+6H_2^2\phi\ln\frac{\phi}{\phi_0}+6H_2^2\phi-2\rho_0
\label{pot}
\ee 
This kind of potential can arise due to higher order loop corrections of the
Coleman-Weinberg potential. 
This type  of potential had been used earlier in quite a few references, 
specially in context of inflationary scenario\cite{BP}. We have plotted the
potential in figure 4. An interesting 
point to note here is that the minimum of the potential has a nonzero value
at which the dilaton stabilises. This asymptotically generates a positive 
cosmological constant which provides a good enough explanation for the missing
energy component. 
We have also plotted in figure 5, the equation of state for the dilaton field,
where one finds that the equations of state stabilises at -1 in late times
suggesting that the dilaton field gives rise to a positive cosmological
constant in late time.

\begin{figure}
\centering
\leavevmode\epsfysize=7cm \epsfbox{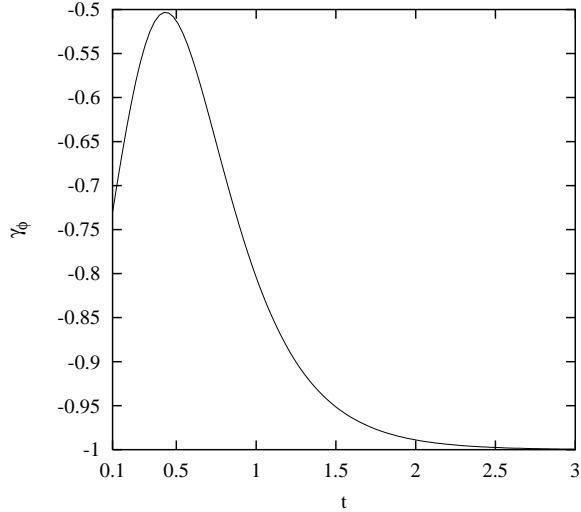}
\caption{Plot of the equation of state for the scalar field 
$\gamma_{\phi}$ {\it vs} $\phi$}
\label{fig5}
\end{figure}
The effective pressure of the CDM fluid is given by
\be
P=\left\{e^{-\alpha(t-t_0)}\left[\frac{5}{8}H_2(16H_1-9\beta)\right]-e^{-2\alpha(t-t_0)}(\beta-3H_1)[\beta(1+\omega)+H_1]\right \}\phi_0Exp[\frac{\beta}{\alpha}e^{-\alpha(t-t_0)}]-\rho_0
\label{press}
\ee   
\begin{figure}
\centering
\leavevmode\epsfysize=7cm \epsfbox{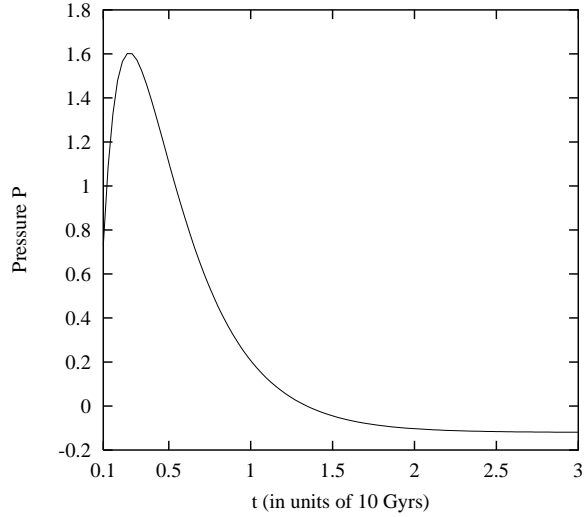}
\caption{Plot of the effective pressure P}
\label{fig6}
\end{figure}
In figure 6 we have plotted the effective pressure $P$ of the CDM. It has
shown that the CDM has very small effective negative pressure in late
time. This is similar to what Chimento et.al\cite{chpav} and also Zimdahl
et.al\cite{zim} assumed in their investigations.

In conclusion, we have studied the cosmological evolution with the low
energy effective string action. We have assumed the form of the dilaton field
in such a way so that it stabilises quickly with time resulting the transition
to the GR theory. Hence the problem of having a higher value of the parameter
$\omega$ consistent with Solar system observation is no more in our theory
which is present in most of the quintessence models in scalar tensor theories
studied in recent times. Moreover the scale factor of the universe in our
model is such that one can have a decelerating universe in early time but in
late time the universe becomes accelerating which is consistent with the
recent supernovae observations. In fact the universe asymptotically becomes a
de-Sitter universe.
We have also calculated the form of potential
that can give rise to such kind behaviour. It has been shown that the
potential has a non zero minima where the dilaton will stabilise and this will
give give rise to a positive cosmological constant asymptotically. The
behaviours of the energy densities of the matter and scalar field show that
although in early time, the energy density of the matter field is
greater than that of the scalar field but in late times the scalar field will
dominate explaining the missing energy component.
We have fixed our initial time at 1 Gyr. But initially the universe is decelerating as well as the matter energy density dominates over that of the
scalar field, hence one can extend our model back in earlier time to meet the
constraints from the  big-bang nucleosynthesis.

It seems that this model can be a viable model, at least theoretically, to
explain the late time acceleration of the universe and the missing energy
component. It will be interesting to see whether this model actually matches
with Type Ia supernovae observations and also with CMBR experiments like
BOOMERANG and MAXIMA.
\vskip 5mm
We gratefully acknowledge Dileep P.Jatkar and Debajyoti Choudhury for useful
 discussions.

\end{document}